\documentclass[acmlarge,manuscript,nonacm]{acmart}

\AtBeginDocument{%
  \providecommand\BibTeX{{%
    \normalfont B\kern-0.5em{\scshape i\kern-0.25em b}\kern-0.8em\TeX}}}

\setcopyright{acmlicensed}
\copyrightyear{2018}
\acmYear{2018}
\acmDOI{XXXXXXX.XXXXXXX}

\begin{document}

\title{Enhancing ICU Patient Recovery: Using LLMs to Assist Nurses in Diary Writing}

\author{S. Kernan Freire}
\orcid{0000-0001-8684-0585}
\affiliation{%
  \institution{Delft University of Technology}
  \streetaddress{Landbergstraat 15}
  \city{Delft}
  \postcode{2628 CE}
  \country{The Netherlands}}
\email{s.kernanfreire@tudelft.nl}

\author{M.M.C. van Mol}
\orcid{0000-0002-0213-6054}
\affiliation{%
  \institution{Erasmuc MC}
  \streetaddress{Dr. Molewaterplein 40}
  \city{Rotterdam}
  \postcode{3015 GD}
  \country{The Netherlands}}
\email{m.vanmol@erasmusmc.nl}

\author{C. Schol}
\orcid{0000-0002-5380-4470} 
\affiliation{%
  \institution{Erasmuc MC}
  \streetaddress{Dr. Molewaterplein 40}
  \city{Rotterdam}
  \postcode{3015 GD}
  \country{The Netherlands}}
\email{c.schol@erasmusmc.nl}

\author{E. Ozcan Vieira}
\orcid{0000-0001-9195-2960}
\affiliation{%
  \institution{Delft University of Technology}
  \streetaddress{Landbergstraat 15}
  \city{Delft}
  \postcode{2628 CE}
  \country{The Netherlands}}
\email{e.ozcan@tudelft.nl}

\renewcommand{\shortauthors}{Kernan Freire, et al.}

\begin{abstract}
Intensive care unit (ICU) patients often develop new health-related problems in their long-term recovery. Health care professionals keeping a diary of a patient’s stay is a proven strategy to tackle this but faces several adoption barriers, such as lack of time and difficulty in knowing what to write. Large language models (LLMs), with their ability to generate human-like text and adaptability, could solve these challenges. However, realizing this vision involves addressing several socio-technical and practical research challenges. This paper discusses these challenges and proposes future research directions to utilize the potential of LLMs in ICU diary writing, ultimately improving the long-term recovery outcomes for ICU patients.
\end{abstract}

\maketitle


\section{Introduction and Background}
Advanced treatments and sophisticated technological interventions in critical care medicine have significantly increased the survival rates of patients in the intensive care unit (ICU). Despite this progress, patients often face various health-related challenges in their long-term recovery~\cite{inoue_post-intensive_2019,herridge_outcomes_2023}. More than half of patients develop new physical, psychological, and/or cognitive problems following their ICU admission~\cite{geense_new_2021}, collectively referred to as  Post Intensive Care Syndrome (PICS)~\cite{davidson_post-intensive_2013,svenningsen_post-icu_2017}. 

Family members also experience a stressful period, potentially leading to psychological problems addressed as PICS-Family (PICS-F)~\cite{cameraon2016one}. Patient and family-centered care (PFCC) at the ICU, including emotional support and follow-up service, could mitigate the symptoms associated with both PICS and PICS-F. In this study, we explored how an emerging technology, i.e., large language models, could support the emotional well-being of people exposed to critical care.

\subsection{Emotional recovery following ICU admission, the use of diaries}
ICU diaries can be used as part of the PFCC approach, focusing on decreasing symptoms of PICS/PICS-F. These diaries are written in everyday language by healthcare professionals, mostly nurses and family members. Using ICU diaries offers many benefits to all stakeholders involved in the ICU. They contain daily entries detailing the current patient status and descriptions of situations and surroundings. Reading a diary after hospitalization is an effective way of coping with the traumatic aftermath of critical illness, consequently helping to prevent the development of psychological problems ~\cite{mcilroy_effect_2019,barreto_impact_2019}. For family members, the diary provides an opportunity to actively care for their loved ones, thus diminishing feelings of powerlessness and reducing psychological problems~\cite{nielsen_how_2016,schofield_experience_2021}. Digital diaries were developed during the COVID-19 pandemic, receiving positive assessments from family members~\cite{van_mol_usability_2023}. However, implementing digital diaries face some barriers~\cite{haakma_experiences_2022,galazzi_thematic_2023} as will be explained further.

\subsection{Digital diary in the ICU; Barriers in implementation}
High-quality PFCC should be considered a fundamental skill for ICU healthcare professionals~\cite{kang_being_2023}. However, barriers have been identified, hindering the effective implementation of digital ICU diaries. The main barrier is a lack of time for healthcare professionals~\cite{nydahl_how_2014}. Another barrier is the lack of knowledge among nurses regarding what and how to write in the diary~\cite{haakma_experiences_2022}. A potential solution to address these challenges could be the integration of Artificial Intelligence (AI) in the writing process of healthcare professionals. 

\subsection{Large Language Models in Healthcare}
Natural language processing (NLP) is a machine learning technique that involves the processing and analysis of text. Large language models (LLMs), such as ChatGPT, are very effective at generating human-like text. LLMs mark a significant step forward from their predecessors, such as recurrent neural networks (RNNs). Unlike RNNs that process text sequentially and often struggle with long-range dependencies, LLMs can analyze and generate text in parallel, handling extensive context and complex language patterns efficiently~\cite{min2023recent, jawahar-etal-2019-bert, wei2022emergent, zhao2023survey}. These characteristics make LLMs effective writing partners, helping humans by generating outlines, refining text, and adapting it to the reader.

In healthcare, applications of NLP range from supporting triage by analyzing prior medical notes to answering patients' questions as a chatbot~\cite{locke_natural_2021}. Recently, LLMs have received widespread attention in healthcare, leading to the creation of health-specific models, such as Med-Palm~\cite{singhal_large_2023}. Applications include supporting clinical workflow, for example, by generating discharge notes, extracting ecological phenotypes~\cite{fink_potential_2023}, and making medical texts more understandable and empathetic for patients~\cite{li_chatgpt_2024}. These capabilities could help tackle challenges nurses face when using digital diaries in the ICU, however, this remains unexplored in the literature~\cite{li_chatgpt_2024,sallam_chatgpt_2023}.

\section{Towards ICU Diaries Supported by Large Language Models}
ICU diaries can provide a more personable timeline for the ICU admission beyond the medical notes that medical professionals already record.  Our vision is that an LLM-powered tool can support the writing process for nurses, making it more efficient without losing the personal touch of a human writer. In the following sections, we describe this vision in more detail and the associated research challenges.

\subsection{Future Vision}
We envision a collaborative writing process that evolves as nurses become more familiar with the LLM-powered tool's capabilities, and in turn, the tool ``learns'' the nurse's writing style. To begin with, nurses may be unfamiliar with the diary writing process for ICU patients. As such, the tool can help nurses figure out what and how to write. At this stage, the tool asks for key information about the situation and generates an example diary entry for the nurse. As the nurse becomes more familiar with the process and expectations, they can start adjusting the entries themselves or write them from scratch. At this stage, the tool can provide in-text suggestions on how to write empathetically and understandably for the patients. Over time, the tool will amass a database of entries about individual patients written by the nurse, allowing the tool to align with their writing style~\cite{min2023recent}. In turn, the nurse can enter a few keywords to generate a diary entry, saving time. Thus, this collaborative process allows for growth and adaptation both on the human and technological sides. 

The tool must support various diary entry themes and modalities, primarily text and images. Prior work by~\citet{galazzi_thematic_2023} has shown that ICU diary entries fall under four main themes with ten sub-themes, namely, Presenting (Places and people; Diary project), Intensive Care Unit Stay (Clinical events; What the patient does; Patient support), Outside the Hospital (Family and topical events; The weather), Feelings and Thoughts (Encouragement and wishes; Farewell; Considerations). While information about patients’ specific ICU experience can only be filled by attending nurses and families, non-patient-related topics, such as the weather and national events, are publicly available via application programming interfaces (APIs). When relevant, the tool could use APIs to enrich the diary entries. Similarly, the tool could access recent entries to the medical documentation records or visitor calendar; however, this poses several technical and ethical challenges as the implications of using LLMs for personal information exchange are not fully understood.

\subsection{Research Challenges}
To realize the vision described above, several socio-technical research challenges must be tackled. These research challenges are discussed under four predefined themes, inspired by prior work that evaluated novel digital tools in support of nurse training~\cite{martinez2023lessons}.

\textit{Space and place} To better understand the tool's impact on the existing ICU environment, research could empirically evaluate how an LLM-based diary task reduces nurses’ daily workload and how it allows for more humanized care.

\textit{Technology} The technological research challenges revolve around optimizing the LLMs for the tool. Important considerations include using specialized LLMs for different aspects, such as suggestion topics, making entries more empathetic, generating entries based on keywords, and querying APIs. Indeed, the desired tool behavior could be realized by combining several techniques, including fine-tuning models~\cite{dingParameterefficientFinetuningLargescale2023}, reasoning strategies (e.g., self-discovery~\cite{zhouSelfDiscoverLargeLanguage2024a}), and/or retrieval augmented generation~\cite{lewis2020retrieval}. Several technical limitations may also be imposed due to the ethical concerns of storing and sharing the patients' personal data, for example, the necessity to host LLMs locally in hospitals or at secure providers. Furthermore, it is imperative to identify the risks associated with model hallucinations and bias and how these can be mitigated~\cite{sallam_chatgpt_2023}.

\textit{Design} The tool's design will play a key role in its usability and adoption. Therefore, research should explore options for interaction modalities, how the tool's outputs are displayed (e.g., in-text suggestions), and integrated into the nurses' workflows.

\textit{Social factors} As with the introduction of any new tool, the end users will likely have several concerns and expectations that should be addressed as early as possible. Therefore, it is important to conduct preliminary research, such as semi-structured interviews and technology probes, to elicit any concerns and requirements from the nurses, patients, and other stakeholders. This will make the tool more socially responsible and human-centered~\cite{shneiderman2022human}.



\section{Conclusion}
Large language models have become increasingly popular in the past few years, especially in non-critical contexts. In this study, we explored the potential of LLMs in supporting ICU nurses in writing diary entries for critically ill patients. While this novel technology has much to offer to humanize a highly technological environment, embedding it in nurses' routines seamlessly, LLM-powered tools must be explored thoroughly to understand the socio-technical limitations, opportunities, and risks.



\begin{acks}
We acknowledge Kaixin Tang for her graduation project, which inspired this work.
\end{acks}

\bibliographystyle{ACM-Reference-Format}
\bibliography{references}



\end{document}